\documentclass[11pt]{article}

\def\ve{\varepsilon}
\def\bfrac#1#2{\left(\frac{#1}{#2}\right)}
\def\mone{$^{-1}$}
\def\p#1{$^{#1}$}
\renewcommand{\thefootnote}{\fnsymbol{footnote}}

\begin{document}
%%%%%%%%%%%%%%%%%%%%%%%%%%%%%%%%%%%%%%%%%%%%%%%%

\begin{titlepage}

%{\flushright{hep-th/0310015\\}}

\vspace*{1cm}
\begin{center}
{\bf \LARGE Vacuum Dominance and Holography}
\end{center}

\vspace{1cm}
\begin{center}
{\sc T. R. Mongan \footnote{E-mail: tmongan@mail.com}} \\
{\it 84 Marin Avenue\\
Sausalito, CA 94965, U.S.A.}
\end{center}

\vspace{1cm}

\begin{abstract}
\noindent
A cosmological event horizon develops in a vacuum-dominated Friedmann
universe.  The Schwarschild radius of the vacuum energy within the
horizon equals the horizon radius.  Black hole thermodynamics and the
holographic conjecture indicate a finite number of degrees of freedom
within the horizon.  The average energy per degree of freedom equals
the energy of a massless quantum with wavelength of the horizon
circumference.  This suggests identifying the degrees of freedom with
the presence or absence, in each Planck area on one horizon quadrant,
of a $_0S_2$ vibrational mode of the horizon with a $z$ axis passing
through that area.  Pressure waves on the horizon (the superposition
of $_0S_2$ vibrational modes) can be envisioned to propagate into the
observable universe within the horizon at the speed of light.  So, the
vacuum energy and pressure throughout the observable universe could
(in principle) be determined from the vacuum equation of state.

\vspace{1cm}
\end{abstract}
\end{titlepage}
\renewcommand{\thefootnote}{\arabic{footnote}}
\setcounter{footnote}{0}

\section*{A vacuum-dominated universe is asymptotic to de Sitter space}
Friedmann's equation for the radius of curvature $R$ of a closed homogeneous 
isotropic universe is
\[
\dot{R}^2 - \bfrac{8\pi G}{3}
\left[ \ve_r \bfrac{R_0}{R}^4 + \ve_m \bfrac{R_0}{R}^3 + \ve_v\right]
\bfrac{R}{c}^2= - c^2
\]
where $\ve_r$, $\ve_m, \ve_v$ and $R_0$  are, respectively, the present values 
of the radiation, matter and vacuum energy densities and the radius of 
curvature.   

Astrophysical measurements indicate the expansion of the universe is 
accelerating, and the energy density of the universe is dominated by a 
cosmological constant/vacuum energy density with $\Omega_\Lambda=0.7$\,.
The cosmological constant is  
related to the vacuum energy density by $\Lambda = 8\pi G \ve_v/c^4$.
At late times,  $R\rightarrow \infty$ and the radiation and  
matter energy density (and the curvature energy) are driven to zero by the 
expansion of the universe.  The Friedmann equation becomes
\[
\bfrac{\dot{R}}{R}^2 = \frac{8\pi G \ve_v}{3c^2} = \frac{\Lambda c^2}{3}.
\]
  So, our universe 
is asymptotic to a de Sitter space (the vacuum solution to Einstein's equations 
with a positive cosmological constant), and the asymptotic value of the Hubble 
constant is $H = \dot{R}/R=c \sqrt{\Lambda/3}$.
 
\section*{Horizon radius equals Schwarschild radius of vacuum energy
  within horizon} There is a cosmological event horizon in a de Sitter 
space, with horizon radius $R_H=c/H=\sqrt{3/\Lambda}$ (see, e.g,
\cite{Carniero, Padman, Frolov}).  Taking the Hubble constant as $H_0
= 65$ km~sec\mone~Mpc\mone, the critical density $\rho_c =
\frac{3H_0^2}{8\pi G} = 7.9 \times 10^{-30}$~g~cm\p{-3}, the vacuum energy 
density $\ve_v= 0.7 \rho_c\, c^2=5.0 \times
10^{-9}$~g~cm\p{2}~sec\p{-2}~cm\p{-3}, and  
$\Lambda = 1.0 \times 10^{-56}$ cm\p{-2}.  

The total vacuum energy within the horizon radius $R_H$ is 
$\ve_v \frac{4\pi}{3}\bfrac{3}{\Lambda}^{3/2} =
\frac{c^4}{2G}\sqrt{\frac{3}{\Lambda}}$.  
The mass
equivalent of this energy, $M=\frac{c^2}{2G}\sqrt{\frac{3}{\Lambda}}$,
is the mass of the observable part of the 
vacuum-dominated universe (analogous to Wesson's ``Einstein mass" \cite{Wesson}).

The Schwarschild radius $R_S$ of a mass $M$ is $R_S = 2GM/c^2$, so the
Schwarschild radius 
of the vacuum energy within the horizon is $R_S = \sqrt{3/\Lambda} =
R_H$.  Therefore, the volume 
within the horizon radius $R_H$ is analogous to the inside of a black
hole with radius $R_S =\sqrt{3/\Lambda} = R_H$.

The relation $\frac{M}{R}\approx \frac{c^2}{2G}$, where $M$ and $R$
are the mass and 
radius of the universe, is sometimes obtained by invoking Mach's
principle \cite{Funkhouser,Tripathy}.  Because the mass equivalent of the vacuum energy
within the horizon is $M=\frac{c^2}{2G}\sqrt{\frac{3}{\Lambda}}$,
the relation $\frac{M}{R} \approx \frac{c^2}{2G}$ holds whenever the
radius of a 
vacuum-dominated Friedmann universe approaches or exceeds the horizon
radius and is approximately true today.
 
\section*{Finite degrees of freedom within the horizon}

The area of the cosmological horizon at late times in a
vacuum-dominated Friedmann universe is $A=4\pi R_H^{2}=12 \pi/\Lambda$.
The holographic conjecture then indicates the number of observable
degrees of freedom in the universe is $N=A/4=\pi R_H^2 =
3\pi/\Lambda$, where $A$ is measured in Planck units.  Thus, $N=\pi
R_H^2 / \delta^2 = 3\pi/(\Lambda \delta^2)$, where $\delta$ is the
Planck length.

Black hole thermodynamics alone indicates the number of observable
degrees of freedom in a black hole with surface area $A$ is $N=A/4$,
where $A$ is measured in Planck units.  So, considering black holes
within the horizon with radii approaching the horizon radius, the
above result from the holographic conjecture is not surprising,
 
\section*{Average energy per degree of freedom}

For {\it any} Schwarschild black hole, with $M=\frac{c^2 R_S}{2G}$,
the number of degrees of freedom is $N=A/4=\pi R_S^2/\delta^2$.  
Defining the Planck length by $\delta=\sqrt{\frac{hG}{c^3}}$, instead of the
usual $\delta=\sqrt{\frac{\hbar G}{c^3}}$, the average energy per  
degree of freedom is  $\frac{E}{N}=\frac{c^4 \delta^2}{2\pi G R_S} =
\frac{h c}{2\pi R_S}$.

Considering the volume of a vacuum-dominated universe within the
horizon, the total vacuum energy within the horizon is
$\frac{c^4}{2G}\sqrt{\frac{3}{\Lambda}} = \frac{c^4 R_H}{2G}$, and the
number of degrees of freedom is $N=\frac{3\pi}{\Lambda \delta^2} = \pi
\frac{R_H^2}{\delta^2}$.  So, the average energy per degree of freedom
in a vacuum-dominated Friedmann universe at late times is $\frac{c^4
  \delta^2}{2\pi G R_H}= \frac{hc}{2\pi R_H} = \frac{\hbar c}{R_H}$.
Since $\hbar c= 197.32$ MeV~Fermi, the average energy per degree of
freedom in the observable vacuum-dominated universe at late times is
about $10^{-33}$ electron volts.  The mass equivalent of this energy
is approximately the quantum of mass (about $2 \times 10^{-65}$ g)
identified by Wesson \cite{Wesson}.

The energy of a massless quantum with wavelength $\lambda$ is
$hc/\lambda$.  So, if the energy per  
degree of freedom in the observable vacuum-dominated universe at late times is 
carried by massless quanta, the average wavelength of those quanta is
$2\pi R_H = 2\pi \sqrt{3/\Lambda} = 1.09 \times 10^{29}$~cm, the
circumference of the horizon.  The average frequency of these  
quanta is $c/\lambda= 2.75 \times 10^{-19}/ \mbox{\,sec} = 8.67 /
10^{12} \mbox{\,yr}$. 

The surface temperature of a Schwarschild black hole of mass $M$ is $T
=\frac{\hbar c^3}{8\pi G M k} = \frac{\hbar c}{4\pi R_S k}$.
A Schwarschild black hole with solar mass $2 \times 10^{33}$~g has a surface
temperature of $6 \times 10^{-8}{\ }^\circ\mbox{K}$, so the surface of any
large Schwarschild 
black hole is a low temperature environment.  In particular, the
surface temperature of a Schwarschild black hole with radius 
$R_S = \sqrt{3/\Lambda} = R_H$ is $T=\frac{\hbar c}{4\pi R_H k} =
10^{-30} {\ }^\circ\mbox{K}$.
 
\section*{Possible connection to holography}

The holographic conjecture and black hole thermodynamics suggests the
number of bits of information necessary to describe the observable
universe of radius $R_H$ within the horizon is
$N=\pi\frac{R_H^2}{\delta^2}$.  This is the number of ``pixels" of area
$\delta^2$ on the surface of one quadrant of the horizon.

The lowest frequency (lowest energy) vibrational mode of a spherical
surface is the $_0S_2$ ``football" mode \cite{Stacey} where the sphere is
alternately elongated and compressed along its $z$ axis.  The pressure
waveform corresponding to this mode, on the circumference in any plane
including the $z$ axis, has wavelength $\pi R$, where $R$ is the
radius of the sphere.  A massless quantum with wavelength $\pi R_H$
has twice the energy of a massless quantum with wavelength $2\pi R_H$
and thus twice the average energy per degree of freedom in the
observable vacuum-dominated universe at late times.  Because the
surface temperature on the horizon is $10^{-30}{\ }^\circ\mbox{K}$,
only the lowest energy excitations should occur on the surface, if the
excitations have integral spin.

Suppose each pixel (bit) on one quadrant of the horizon has value 1 if
the $z$ axis of a lowest frequency vibrational mode lies within that
pixel and zero (no excitation) otherwise.  If values 0 and 1 are
randomly assigned to each pixel, $N/2$ lowest frequency modes will be
excited, and the average energy per degree of freedom will be
$\frac{hc}{2\pi R_H}$.  Because of the symmetry of the lowest
frequency vibrational mode $_0S_2$, specifying the bits on one
quadrant of the horizon determines the pressure waveform
(superposition of $N/2$ lowest frequency modes) on the entire horizon.

The pressure pattern on the horizon (the superposition of the $N/2$
lowest frequency modes excited) can be envisioned to propagate into
the universe within the horizon at the speed of light.  Because the
equation of state of the vacuum energy is known, a gigantic
calculation (possible in principle) could determine the vacuum energy
and pressure throughout the observable universe.  

The scalar field
responsible for the vacuum energy might be related to the size of 
compact extra dimensions, as suggested by Turner \cite{Turner} and
considered in \cite{Mongan}.  If so, the vacuum energy and pressure at any
point could be  
related to perturbations in the size of the compact dimensions at that
point.  Since distortions of compact dimensions are seen in string/M
theory as fermions and bosons, wave patterns on the horizon might
suffice to reconstruct the distribution of matter and energy within
the observable universe.  This approach could only be applied to
Schwarschild black holes within our universe if the equation of state
for the energy inside those black holes could be determined.  Although
this discussion focuses on closed vacuum-dominated Friedmann
universes, the argument can be extended to flat and open
vacuum-dominated Friedmann universes.

%%%%%%%%%%%%%%%%%%%%%%%%%%%%%%%%%%%%%%%%%%%%%%%%%%%%%%%%%%%%%%%%%%%%%%
\end{document}